\documentclass[%
 reprint,
 amsmath,amssymb,
 aps,
prb,
floatfix
]{revtex4-2}

\usepackage{graphicx}
\usepackage{bm}

\newcommand\ee{\mathrm{e}}
\newcommand\cc{\mathrm{c}}

\newcommand\st{^\mathrm{st}}
\newcommand\tth{^\mathrm{th}}
\newcommand\nm{\,\mathrm{nm}}

\newcommand\ii{\mathrm{i}}

\newcommand\omegap{\omega_\mathrm{p}}

\renewcommand\Re{\mathop{\rm Re}\limits}
\renewcommand\Im{\mathop{\rm Im}\limits}
\newcommand\lt\left
\newcommand\rt\right

\newcommand\mat[1]{#1}

\begin{document}
\title{Coupled-wave formalism for bound states in the continuum \\in guided-mode resonant gratings}

\author{Dmitry A. Bykov}
 \email{bykovd@gmail.com}
\author{Evgeni A. Bezus}
\author{Leonid L. Doskolovich}

\affiliation{Image Processing Systems Institute --- Branch of the Federal Scientific Research Centre ``Crystallography and Photonics'' of Russian Academy of Sciences, 151 Molodogvardeyskaya st., Samara 443001, Russia}
\affiliation{Samara National Research University, 34 Moskovskoye shosse, Samara 443086, Russia}

\date{\today}

\begin{abstract}
We present simple yet extremely accurate coupled-wave models describing
the formation of bound states in the continuum (BICs) in 1D-periodic
guided-mode resonant gratings (GMRGs) consisting of a slab waveguide
layer with a binary grating attached to one or both of its
interfaces. Using these models, we obtain simple closed-form
expressions predicting the locations of the BICs and quasi--BICs in the $\omega$--$k_x$
parameter space. We study two mechanisms of the BIC formation:
coupling between two counter-propagating guided modes and
coupling between a guided mode and a Fabry--P{\'e}rot mode. The BIC
conditions for the two considered mechanisms are formulated in terms
of the scattering coefficients of the binary grating. The predictions
of the presented models are in excellent agreement with the results of
rigorous numerical simulations obtained using the Fourier modal method.
\end{abstract}

\maketitle

\section{\label{sec:introduction}Introduction}

In recent years, investigation of bound states in the continuum (BICs) has attracted a lot of attention. BICs were for the first time theoretically predicted in 1929 by von Neumann and Wigner for an electronic system with an artificially tailored potential~\cite{NW}. In the last decade, BICs were discovered and studied in various photonic structures (see the review paper~\cite{HsuReview} and references therein). In contrast to the conventional bound states, BICs are the eigenmodes that, although supported by a structure with open scattering channels, remain perfectly confined (i.\,e.\ have an infinite lifetime and an infinite quality factor). There are different mechanisms preventing the leakage of the mode energy to the open channels, including symmetry mismatch~\cite{HsuReview} and destructive interference of resonances~\cite{FW}. 

Photonic BICs were studied in various periodic structures (diffraction gratings, photonic crystal slabs, and infinite arrays of dielectric rods or spheres)~\cite{Marinica, Hsu, Bulgakov:2018:josab, Azzam:2018:prl, Bulgakov:2018:pra, Sadreev2, Blanchard, Shipman, Sadrieva, Sadrieva2, my:Bykov:2015:pra, Yoon:2015:sr},
in defects and interfaces of photonic crystals~\cite{Hsu2, Sadreev1, Sadreev3},
in arrays of optical waveguides~\cite{A1,A2,A3}, 
and in photonic rib waveguides~\cite{Zou, my:Bezus:2018:pr},
among others~\cite{HsuReview}.
The phenomenon of BIC is not only of great theoretical interest, but has a wide range of practical applications, since a small perturbation in the parameters of the structure or of the incident radiation leads to the collapse of a BIC to a Fano resonance with an extremely high quality factor.
The design of such resonators is important, in particular, for the development of lasers, sensors, and filters.

Periodic optical structures such as diffraction gratings or photonic crystal slabs (PCSs) constitute one of the basic building blocks of photonics.
For these structures, advanced theoretical models and efficient numerical methods giving a deep insight into their optical properties have been developed.
The interest in diffraction gratings and PCSs is due to the fact that they exhibit a wide range of optical phenomena.
In particular, most of the known \emph{resonant} effects arising in optics and photonics can be investigated by studying diffraction gratings.
It is, thus, not a surprise that starting from the pioneering paper~\cite{Marinica}, a vast majority of the published papers studying photonic BICs focuses on gratings and other periodic structures~\cite{Marinica, Hsu, Bulgakov:2018:josab, Azzam:2018:prl, Bulgakov:2018:pra, Sadreev2, Blanchard, Shipman, Sadrieva, Sadrieva2, my:Bykov:2015:pra, Yoon:2015:sr}.
In Ref.~\cite{Marinica}, BICs arising due to Fabry--P{\'e}rot interference between a pair of resonant gratings were theoretically and numerically studied.
Extensive numerical~\cite{Sadreev2, Sadrieva, Yoon:2015:sr, Bulgakov:2018:josab} and experimental~\cite{Sadrieva2} investigations of BICs in structures with 1D periodicity were carried out.
In particular, field enhancement effects near BICs were discussed in Ref.~\cite{Yoon:2015:sr}.
In Ref.~\cite{Hsu}, an experimental study of a PCS with 2D periodicity was presented.
A recent work~\cite{Azzam:2018:prl} demonstrated the existence of quasi-BICs resulting from the coupling of waveguide and plasmonic modes in a metallic grating located on a slab waveguide.
Simple approximate models for the BICs based on power series expansion were proposed in papers~\cite{Shipman, my:Bykov:2015:pra, Blanchard}.
In particular, in~\cite{Shipman, my:Bykov:2015:pra}, the case of symmetry-protected BICs in PCS with 1D periodicity was considered.
In a recent paper~\cite{Bulgakov:2018:pra}, BICs in low-contrast PCSs were theoretically described using Fourier series expansion of the dielectric permittivity.

\begin{figure}[tb]
	\centering
		\includegraphics{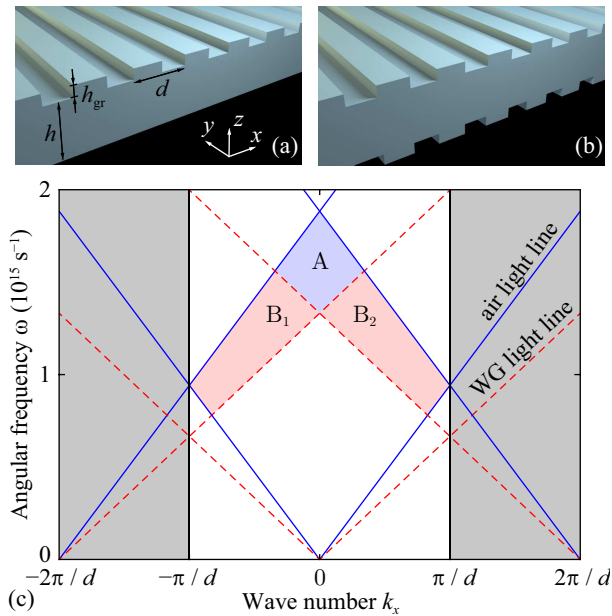}
	\caption{\label{fig:structure}Guided-mode resonant gratings consisting	of a slab waveguide layer with a binary grating attached to one~(a) or both~(b) of its interfaces. 
	Light lines of $\pm 1^{\rm st}$ and $0^{\rm th}$ diffraction orders in the substrate and superstrate (solid blue lines) and in the waveguide layer (dashed red lines) (c)}
\end{figure}

In this paper, we investigate the BICs and quasi--BICs supported by binary gratings located on a slab waveguide surface [Fig.~\ref{fig:structure}(a),~(b)]. 
We will refer to these structures as guided-mode resonant gratings (GMRGs).
The eigenmodes of the GMRGs considered in the present work arise due to constructive or destructive interference of the plane waves inside the waveguide layer.
This allows us to formulate simple coupled-wave models describing the optical properties of the GMRGs.
The developed models do not utilize series expansion techniques, which makes them extremely accurate.
Using these models, we derive simple closed-form expressions predicting the locations of the BICs in the $\omega$--$k_x$ parameter space.
We investigate two different mechanisms of the BIC formation: coupling of two counter-propagating waveguide modes, 
and coupling of a waveguide mode with a Fabry--P{\'e}rot mode.

The paper is organized in five sections.
Following the Introduction, Section~\ref{sec:geometry} describes the geometry of the considered structure and revisits some basic aspects of the grating theory.
In Section~\ref{sec:a}, we study BICs arising due to coupling of two counter-propagating waveguide modes, which happen near the center of the first Brillouin zone.
In Section~\ref{sec:b}, we study BICs caused by coupling of a waveguide mode with a Fabry--P{\'e}rot mode.
Section~\ref{sec:conclusion} concludes the paper.

\section{\label{sec:geometry}GMRG geometry}
 
Let us start with a guided-mode resonant grating (GMRG) shown in Fig.~\ref{fig:structure}(a).
The considered GMRG consists of a binary grating with period $d$ placed on top of a slab waveguide (WG) layer with thickness $h$ and refractive index $n$.
The binary grating is assumed to be non-resonant, i.\,e.\ all the resonances (and the eigenmodes) of the GMRG arise due to multiple reflections of waves inside the WG layer.
The eigenmodes can be excited by an incident plane wave, which is defined by the angular frequency $\omega$ and the in-plane wave vector component $k_x = k_0 \sin\theta$, where $\theta$ is the angle of incidence and $k_0 = \omega/\cc$ is the free-space wave number.
Let us note that in the present work we consider only the case of planar diffraction ($k_y = 0$).

We will assume that the grating is subwavelength, i.\,e.\ only the $0\tth$ diffraction orders exist in the substrate and superstrate regions.
However, in the WG layer, which has a higher refractive index comparing to the substrate (superstrate), several diffraction orders may propagate.
For the $m\tth$ diffraction order inside the WG layer, the $x$-component of the wave vector has the following form:
$$
k_{x,m} = k_x + \frac{2\pi}{d}m.
$$
The $z$-component of the wave vector reads as
\begin{equation}
\label{kz}
k_{z,m} = \sqrt{k_0^2 n^2 - k_{x,m}^2}
\end{equation}
and is real for the propagating diffraction orders and imaginary for the evanescent ones.

Figure~\ref{fig:structure}(c) shows the light lines (light cones) of the $0\tth$ and $\pm1\st$ diffraction orders in the substrate (superstrate) and in the WG layer.
In the shaded regions marked as $A$ and $B_{1,2}$, the substrate and superstrate support only the $0\tth$ diffraction orders, whereas in the WG layer, also the $\pm1\st$ diffraction orders propagate. 
In region $A$, both $-1\st$ and $+1\st$ orders exist in the waveguide layer, while in the regions $B_{1,2}$ only one of these diffraction orders is present.
The regions $A$ and $B_{1,2}$ are the regions where GMRGs exhibit pronounced resonant properties, 
which allow one to use GMRGs as optical filters known as guided-mode resonant filters.

In the following Section~\ref{sec:a}, we consider BICs supported by the GMRG operating in the region $A$.
In Section~\ref{sec:b}, we will focus on the regions $B_{1,2}$.


\section{BIC and quasi-BIC emerging from coupling of two waveguide modes}\label{sec:a}

\subsection{\label{ssec:cwt1}Coupled-wave model}
In this subsection, we present a coupled-wave model of the GMRG operating in the region~$A$ [see Fig.~~\ref{fig:structure}(c)]. 
This model allows us to obtain simple expressions for the transmission and reflection spectra of the GMRG.
As we show below, analysis of these expressions enables predicting the positions of the BICs and quasi-BICs in the $\omega$--$k_x$ plane.

\begin{figure}
	\centering
		\includegraphics{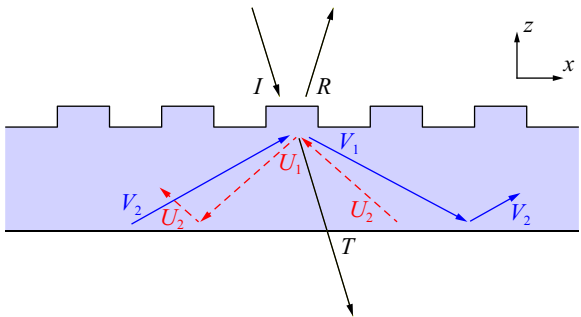}
	\caption{\label{fig:str1}Propagating plane waves inside the GMRG considered in the coupled-wave model.}
\end{figure}

To formulate the coupled-wave model, we consider only the propagating diffraction orders (plane waves) in the superstrate, substrate, and in the WG layer of the GMRG.
In doing so, we assume that the WG layer is thick enough so that the near-field interactions between its interfaces caused by the evanescent diffraction orders can be neglected.
The considered propagating plane waves are shown with arrows in Fig.~\ref{fig:str1}.
The field in the superstrate comprises two plane waves: the incident wave $I$ and the reflected wave $R$.
The field in the WG layer is represented by four plane waves: the $\pm 1\st$ diffraction orders ($U_1$ and $V_1$) and the same plane waves after total internal reflection from the lower interface of the WG layer ($U_2$ and $V_2$).
The two $V$-waves form the rightward-propagating mode of the waveguide.
Similarly, the two $U$-waves form the waveguide mode propagating to the left.
We assume that the $0\tth$ transmitted diffraction  order of the grating goes through the lower interface of the WG layer without reflection (Fig.~\ref{fig:str1}), which gives the transmitted wave $T$.
We will discuss the consequences of this assumption later in Subsection~\ref{ssec:discuss1}.

The letters in Fig.~\ref{fig:str1} denote the complex amplitudes of the considered plane waves.
The amplitudes of the incident, reflected, and transmitted plane waves ($I$, $R$, and $T$) are defined at the upper interface of the grating.
The amplitudes of the upward-propagating waves inside the waveguide ($U_2$ and $V_2$) are defined at the lower interface of the WG layer, whereas the amplitudes of the downward-propagating plane waves ($U_1$ and $V_1$) are defined at the upper interface of the WG layer.

The amplitudes of the $U$-waves at the upper and lower interfaces differ by $\ee^{\ii \phi}$, where the phase is given by $\phi = h k_{z,-1}$.
According to Eq.~\eqref{kz}, this phase has the form
\begin{equation}
\label{phi1}
\phi = h \sqrt{k_0^2 n^2 - \lt(k_x - \frac{2\pi}{d}\rt)^2}.
\end{equation}
Similarly, the complex amplitudes of the $V$-waves change by $\ee^{\ii \psi}$ after propagating between the upper and lower interfaces of the waveguide layer.
The phase $\psi$ is given by
\begin{equation}
\label{psi1}
\psi = h \sqrt{k_0^2 n^2 - \lt(k_x + \frac{2\pi}{d}\rt)^2}.
\end{equation}

The plane waves shown in Fig.~\ref{fig:str1} are coupled by the diffraction grating, which can be described by a $4 \times 4$ scattering matrix $\mat{S}$:
\begin{equation}
\label{S1a}
\begin{bmatrix}
R \\ T \\ U_1 \\ V_1 
\end{bmatrix}
=
\underbrace{
\begin{bmatrix}
r   & t   & d_{ru} & d_{rv} \\
t   & \tilde{r}  & d_{tu} & d_{tv} \\ 
d_{ru} & d_{tu} & r_u & c \\
d_{rv} & d_{tv} & c   & r_v
\end{bmatrix}
}_{\mat{S}}
\begin{bmatrix}
I \\ 0 \\ U_2 \ee^{\ii \phi} \\ V_2 \ee^{\ii \psi}
\end{bmatrix},
\end{equation}
where the zero in the right-hand side means that there is no wave incident from the substrate.
The elements of the scattering matrix are the coupling coefficients having the following notation.
The letters $d$, $r$, and $t$ denote diffraction, reflection, and transmission, respectively.
The subscripts in the diffraction coefficients denote the scattering channels being coupled.
Finally, the coefficient $c$ describes coupling between the $U$- and $V$-waves.

Note that the scattering matrix in Eq.~\eqref{S1a} is symmetric due to reciprocity~\cite{Gippius:2005:prb}, therefore, it contains only 10 unique elements.
Let us further note that for a lossless structure, the scattering matrix is unitary~\cite{Gippius:2005:prb}. 
Therefore, the defined coefficients are not arbitrary but are subject to the energy conservation law.

%

As it is shown in Fig.~\ref{fig:str1}, we assumed that the wave $T$ goes through the lower interface of the waveguide layer without reflection.
However, the waves $U_1$ and $V_1$ undergo total internal reflection at the lower interface. By denoting the corresponding reflection coefficients as $r'_u$ and $r'_v$, we obtain
\begin{equation}
\label{S1b}
\lt\{
\begin{aligned}
U_2 &= r'_u U_1 \ee^{\ii \phi},\\
V_2 &= r'_v V_1 \ee^{\ii \psi}.
\end{aligned}
\rt.
\end{equation}

Let us consider the diffraction of a unity-amplitude incident wave ($I = 1$).
In this case, by solving Eqs.~\eqref{S1a} and~\eqref{S1b}, we obtain the transmission coefficient of the grating:
\begin{widetext}
\begin{equation}
\label{solT1}
T = t + \frac
{d_{rv} d_{tv} \lt(\ee^{-2\ii \phi}/r'_u - r_u \rt) + 
 d_{ru} d_{tu} \lt(\ee^{-2\ii \psi}/r'_v - r_v \rt) + 
c (d_{rv} d_{tu} + d_{ru} d_{tv})
}
{\lt(\ee^{-2\ii \phi}/r'_u - r_u \rt)
 \lt(\ee^{-2\ii \psi}/r'_v - r_v \rt) - c^2 }.
\end{equation}
\end{widetext}
A similar expression can be obtained for the reflection coefficient (complex amplitude of the $0\tth$ reflected diffraction order).

\subsection{Bound states in the continuum}\label{ssec:bic1}
In this subsection, using the approximate Eq.~\eqref{solT1}, we study the BICs supported by the GMRGs in region~$A$ [see Fig.~\ref{fig:structure}(c)].
However, before discussing the BICs, let us first recall some necessary facts from the theory of resonances.

Usually, resonances in the reflection and transmission spectra are described by a complex frequency $\omegap$ of the corresponding eigenmode of the structure~\cite{my:bykov:2013:jlt}.
The complex eigenfrequencies can be found as complex poles of the transmission or reflection coefficients 
[i.\,e.\ by solving the equation $T(\omegap) = \infty$ or $R(\omegap) = \infty$]~\cite{my:bykov:2013:jlt}.
The real part of $\omegap$ gives the mode excitation frequency.
The imaginary part determines the line width of the resonance,
whereas its inverse, referred to as the lifetime of the resonance, describes the decay rate of the mode.

Bound states in the continuum are the modes with an infinite lifetime and, hence, with a real frequency.
Therefore, the denominator in Eq.~\eqref{solT1} vanishes at a real frequency of the BIC.
This seemingly contradicts the energy conservation law: 
the transmission coefficient~\eqref{solT1} would diverge once the denominator vanishes.
This contradiction is easily resolved if the BIC conditions result in the vanishing of the numerator as well~\cite{Blanchard}. 
In this regard, we will use the following approach to find the BICs:
first, we simultaneously equate to zero the numerator and the denominator in Eq.~\eqref{solT1};
then, we verify whether the denominator and numerator vanish at a \emph{real} frequency and a \emph{real} wave number $k_x$.

Let us equate the numerator and the denominator in Eq.~\eqref{solT1} to zero.
By solving the resulting system with respect to $\ee^{\ii\phi}$ and $\ee^{\ii\psi}$, we obtain
\begin{equation}
\label{sol1general}
\begin{aligned}
\ee^{\ii\phi} &= \pm \frac{1}{\sqrt{r'_u \lt( r_u - c\, d_{ru}/d_{rv} \rt)}}, 
\\
\ee^{\ii\psi} &= \pm \frac{1}{\sqrt{r'_v \lt( r_v - c\, d_{rv}/d_{ru}\rt)}},
\end{aligned}
\end{equation}
where the plus and minus signs can be chosen independently for $\ee^{\ii\phi}$ and $\ee^{\ii\psi}$.

If Eqs.~\eqref{sol1general} describe a BIC (i.\,e.\ a real-frequency mode), the phases $\phi$ and $\psi$ should be real [see Eqs.~\eqref{phi1} and~\eqref{psi1}]. 
Therefore, the left- and, hence, the right-hand side of Eqs.~\eqref{sol1general} should lie on the unit circle on the complex plane.
Let us show that that is true in the case of normal incidence of light (at $\theta = 0$).

In the case of normal incidence, the scattering problem is symmetric.
Hence, some of the coupling coefficients coincide, namely, 
$d_{ru} = d_{rv}$, $d_{tu} = d_{tv}$, $r_u = r_v$, and $r'_u = r'_v$.
Besides, the phases $\phi$ and $\psi$ are equal.
In this case, Eqs.~\eqref{sol1general} take the following form:
\begin{equation}
\label{sol1}
\ee^{\ii\phi} = \ee^{\ii\psi} = \pm \frac{1}{\sqrt{r'_u \lt( r_u - c \rt)}}.
\end{equation}

Let us prove that the fraction in the right-hand side of Eq.~\eqref{sol1} lies on the unit circle. The term $r'_u$ is the total-internal-reflection coefficient,
hence $|r'_u| = 1$. 
The second term, $r_u - c$, has unit modulus as well, since it is the eigenvalue of the unitary matrix $\mat{S}$ written for the case of normal incidence.
Thus, the exponents $\ee^{\ii\phi}$ and $\ee^{\ii\psi}$ in Eq.~\eqref{sol1} lie on the unit circle and, consequently, $\phi$ and $\psi$ defined by Eqs.~\eqref{phi1} and~\eqref{psi1} are real.
Therefore, we can obtain closed-form expressions for $\phi$ and $\psi$
by taking the argument of the complex numbers in Eq.~\eqref{sol1}:
\begin{equation}
\label{sol2}
\phi = \psi = \pi m - \frac{1}{2} \arg \lt\{ r'_u \lt( r_u - c \rt) \rt\}. 
\end{equation}
Here, $m$ is an integer.
According to Eqs.~\eqref{phi1}, the phase $\phi$ is a positive number.
Therefore, assuming that the $\mathrm{arg}$ value lies within the interval $[0, 2\pi)$, we restrict $m$ to be a positive integer.

Having derived the expressions for $\phi$ and $\psi$, we can obtain the values of $\omega$ and $k_x$ from Eqs.~\eqref{phi1} and~\eqref{psi1}:
\begin{equation}
\label{bicwkx1}
\begin{aligned}
k_x &= \frac{d}{8\pi h^2}\lt(\phi^2 - \psi^2\rt), \\
\omega &= \frac{\cc}{n}\sqrt{ k_x^2 + \frac{\phi^2 + \psi^2}{2 h^2} + \frac{4\pi^2}{d^2} }.
\end{aligned}
\end{equation}
In the case of normal incidence, $\phi=\psi$, $k_x=0$, and the frequency of the BIC is
\begin{equation}
\label{bicwkx2}
\omega = \frac{\cc}{n}\sqrt{\frac{\phi^2}{h^2} + \frac{4\pi^2}{d^2} }.
\end{equation}

\subsection{\label{ssec:discuss1}Discussion of the model}
In Subsection~\ref{ssec:cwt1}, we made an important assumption when formulating the coupled-wave model: we neglected the reflection of the $0\tth$ diffraction order at the lower interface of the WG layer.
This makes the obtained Eq.~\eqref{solT1} describing the transmission coefficient inaccurate.
Surprisingly, the model describing the BICs presented in Subsection~\ref{ssec:bic1} turns out to be exact!
Indeed, the BICs have zero-amplitude transmitted field in the $0\tth$ diffraction order of the GMRG.
Therefore, the amplitude of the $0\tth$ diffraction order, which is incident at the waveguide lower interface (from inside the WG layer), is zero and, hence, no reflection of the $0\tth$ diffraction order occurs at the WG lower interface.
This makes Eqs.~\eqref{sol2} and~\eqref{bicwkx2} accurate when calculating the BICs.

In the previous subsection, we obtained the BIC condition in the case of normal incidence of light on a structure having a vertical symmetry plane.
Therefore, Eqs.~\eqref{sol2} describe the well-known case of symmetry-protected BICs~\cite{HsuReview, my:Bykov:2015:pra}.
In the case of oblique incidence, however, the right-hand sides of Eqs.~\eqref{sol1general} no longer lie on the unit circle.
As a consequence, the BIC condition is violated in the case of oblique incidence, and the infinite-Q BICs collapse to high-Q resonances (``quasi-BICs'').
As we demonstrate in the next subsection, the $(\omega, k_x)$ positions of these quasi-BICs can still be calculated using Eqs.~\eqref{bicwkx1} with the phases $\phi$ and $\psi$ obtained by taking the argument of Eqs.~\eqref{sol1general}.

\begin{figure*}[bt]
	\centering
		\includegraphics{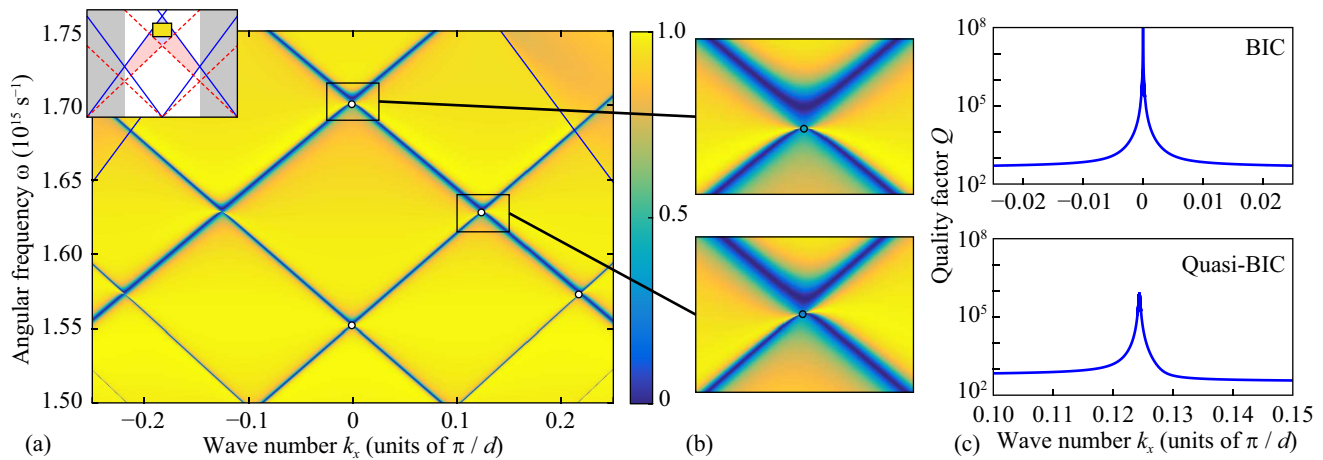}
	\caption{\label{fig:spectrum1}Rigorously calculated transmission spectrum of the considered GMRG~(a) and two magnified fragments~(b) demonstrating the formation of a BIC (upper panel) and of a quasi-BIC (lower panel). The quality factors of the modes corresponding to the magnified fragments~(c).}
\end{figure*}

Let us note that in order to calculate the BIC positions using Eq.~\eqref{bicwkx1} or~\eqref{bicwkx2}, one needs to calculate the phases~$\phi$ and~$\psi$, which depend on the elements of the scattering matrix~$\mat{S}$.
Unfortunately, these elements, being the reflection, transmission, or diffraction coefficients of the grating depend of both the light frequency $\omega$ and wave number~$k_x$.
Therefore, despite its simplicity, Eqs.~\eqref{sol1general},~\eqref{bicwkx1} and Eqs.~\eqref{sol1},~\eqref{bicwkx2} are \emph{nonlinear} equations with respect to $\omega$ and~$k_x$.
Fortunately, these equations can easily be solved iteratively.
We start with some approximate values of $\omega$ and~$k_x$. 
Then, we calculate the scattering matrix $\mat{S}$ and the phases $\phi$ and $\psi$.
Then, we refine the values of~$\omega$ and~$k_x$ using Eqs.~\eqref{bicwkx1} or Eq.~\eqref{bicwkx2}.
Then, we start the next iteration with calculating~$\mat{S}$, and so forth.
This iterative process shows a good convergence.

\subsection{\label{ssec:sim1}Numerical simulations}
Figure~\ref{fig:spectrum1}(a) shows the $\omega$--$k_x$ transmission spectrum of the GMRG with the following parameters:
refractive index $n = 2.1$,
grating period $d = 1000\nm$,
grating height $h_{\rm gr} = 200\nm$,
grating fill-factor $50\%$,
and WG layer thickness $h = 2000\nm$.
No optimization was performed regarding these parameters.
To calculate the transmission spectrum, we used the Fourier modal method~\cite{Moharam:1995:josaa, Li:1996:josaa2}.
The presented spectrum is plotted in the region~$A$ as it is shown in the inset of Fig.~\ref{fig:spectrum1}(a).

Figure~\ref{fig:spectrum1}(b) shows two magnified fragments of the spectrum.
The upper panel shows the BIC region: the resonant line narrows to zero exactly at $k_x = \theta = 0$. 
The lower panel corresponds to the case of oblique incidence: a quasi-BIC is expected according to the model.

In order to distinguish BICs from quasi-BICs, we investigated the quality factor of the resonances $Q = \Re \omegap /(-2\Im\omegap)$ by calculating the complex pole of the scattering matrix~\cite{my:bykov:2013:jlt}.
Figure~\ref{fig:spectrum1}(c) shows the quality factor, which is calculated along the dispersion curves of the resonances shown in Fig.~\ref{fig:spectrum1}(b).
It is evident from Fig.~\ref{fig:spectrum1}(c) that in the case of normal incidence, a BIC is present, while in the oblique incidence case, the quality factor of the resonance is finite, which is the evidence of a quasi-BIC.

To verify the presented coupled-wave model, we calculated the predicted positions of the BICs and quasi-BICs.
The BIC positions were calculated using Eqs.~\eqref{sol2} and~\eqref{bicwkx2}, whereas the quasi-BIC positions were calculated using Eqs.~\eqref{sol1general} and~\eqref{bicwkx1}.
The predicted positions of the BICs and quasi-BICs are shown with black circles in Fig.~\ref{fig:spectrum1}(a) and~\ref{fig:spectrum1}(b).
For illustrative purposes, we show only the positions of the BICs and quasi-BICs having non-negative wave numbers $k_x$.
An excellent agreement between the predicted (quasi-)BIC positions and the corresponding features in the rigorously calculated spectrum confirms the accuracy of the presented coupled-wave model.



\section{\label{sec:b}BIC emerging from coupling of a waveguide mode with a Fabry--P{\'e}rot mode}

In the previous section, we considered the coupling of two counter-propagating waveguide modes excited by $\pm$1-st diffraction orders, neglecting the reflection of the $0\tth$ diffracted order at the lower interface of the WG layer.
However, multiple interference of the $0\tth$ order may by itself result in a resonance, namely, the Fabry--P{\'e}rot resonance.
In this section, we address the question whether the coupling of this ``Fabry--P{\'e}rot mode'' with a waveguide mode can result in a BIC.
We will carry out our analysis for the region $B_2$ [see Fig.~\ref{fig:structure}(c)] (a similar analysis can be performed for the region $B_1$).
In the region~$B_2$, the waveguide layer supports only two propagating diffraction orders, the $0\tth$ and the $-1\st$, while the $+1\st$ diffraction order is evanescent in the WG layer.
Similarly to the previous section, only the $0\tth$ diffraction orders propagate in the substrate and superstrate.

The grating of Fig.~\ref{fig:structure}(a) does not support BICs in the regions $B_{1,2}$ because the Fabry--P{\'e}rot mode leaks out to the substrate via the $0\tth$ diffraction order, as we discussed in Subsection~\ref{ssec:discuss1}.
In order to obtain BICs, we have to change the scattering behavior of the $0\tth$ diffraction order at the lower interface of the structure.
We do this by adding the second diffraction grating (having the same geometry) at the lower interface of the waveguide as shown in Fig.~\ref{fig:structure}(b).
An alternative approach, not considered in this paper, is to increase the refractive index of the superstrate, so that the $0\tth$ diffraction channel in the substrate closes before the one in the superstrate does.


\subsection{\label{ssec:cwt2}Coupled-wave model}

\begin{figure}
	\centering
		\includegraphics{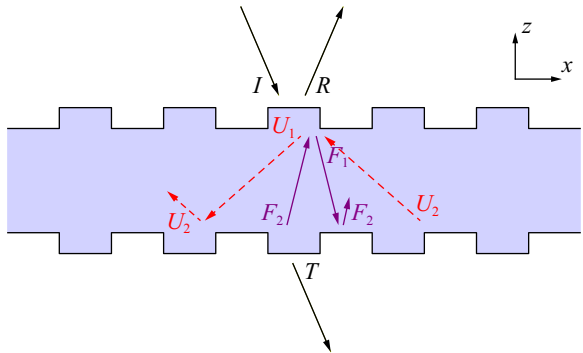}
	\caption{\label{fig:str2}Propagating plane waves inside the GMRG with a horizontal symmetry plane considered in the coupled-wave
model.}
\end{figure}

Similarly to Subsection~\ref{ssec:cwt1}, let us formulate the coupled-wave model for the structure of Fig.~\ref{fig:structure}(b).
To do this, we define the amplitudes of the plane waves as shown in Fig.~\ref{fig:str2}.
Each arrow in Fig.~\ref{fig:str2} denotes a plane wave.
As before, the letters denote the complex amplitudes of the plane waves.
The amplitudes of the incident and reflected plane waves ($I$ and $R$) are defined at the upper interface of the structure, whereas the amplitude of the transmitted wave $T$ is defined at the lower interface.
Inside the waveguide, the amplitudes of the upward-propagating plane waves ($U_2$ and $F_2$) are defined at the lower interface of the WG layer, 
while the amplitudes of the downward-propagating plane waves ($U_1$ and $F_1$) are defined at the upper interface.
The $F$-waves form the Fabry--P{\'e}rot resonance inside the WG layer.

The amplitudes of the $U$-waves at the upper and lower interfaces differ by $\ee^{\ii \phi}$, where the phase $\phi$ is defined by Eq.~\eqref{phi1}.
Similarly, the complex amplitude of the $F$-waves change by $\ee^{\ii \xi}$ when propagating between the upper and lower interfaces of the WG layer.
The phase $\xi$ equals $h k_{z,0}$ and, according to Eq.~\eqref{kz}, reads as
\begin{equation}
\label{xi1}
\xi = h \sqrt{k_0^2 n - k_x^2}.
\end{equation}

In the structure of Fig.~\ref{fig:str2}, the plane waves are coupled by both the upper and lower gratings.
The upper grating can be described by a unitary $3 \times 3$ scattering matrix $\mat{S}$:
\begin{equation}
\label{S2a}
\begin{bmatrix}
R\\F_1\\U_1
\end{bmatrix}
=
\underbrace{
\begin{bmatrix}
r   & t   & d_{ru} \\ 
t   & \tilde{r}  & d_{tu} \\
d_{ru} & d_{tu} & r_u \\
\end{bmatrix}
}_{\mat{S}}
\begin{bmatrix}
I \\ F_2 \ee^{\ii \xi} \\ U_2 \ee^{\ii \phi} 
\end{bmatrix}.
\end{equation}
Here, the coefficients have the same meaning as in Eq.~\eqref{S1a}.
Note that the coefficient $d_{tu}$ in Eq.~\eqref{S2a} describes the coupling between the $0\tth$ and $-1\st$ diffraction orders,
i.\,e.\ between a waveguide mode and a Fabry--P{\'e}rot mode.

Due to the existence of a horizontal symmetry plane, the coupling of plane waves by the lower diffraction grating is described by the very same $3\times 3$ scattering matrix~$\mat{S}$ of Eq.~\eqref{S2a}:
\begin{equation}
\label{S2b}
\begin{bmatrix}
T\\ F_2\\U_2
\end{bmatrix}
=
\mat{S}
\begin{bmatrix}
0\\F_1 \ee^{\ii \xi}\\
U_1 \ee^{\ii \phi} 
\end{bmatrix}.
\end{equation}
Here, zero means that there is no wave incident from the substrate region.

We can now solve Eqs.~\eqref{S2a} and~\eqref{S2b} with respect to~$T$, assuming $I = 1$:
\begin{widetext}
\begin{equation}
\label{solT2}
T = \frac
{\ee^{-\ii\phi-\ii\xi} (d_{ru}^2 \ee^{-\ii\xi} + t^2 \ee^{-\ii\phi}) 
- (d_{ru} \tilde{r} - d_{tu} t)^2 \ee^{-\ii\phi} 
- (d_{ru} d_{tu} - r_u t)^2 \ee^{-\ii\xi}}
{\left[(\ee^{-\ii\xi} - \tilde{r})(\ee^{-\ii\phi} - r_u) - d_{tu}^2\right]
\left[(\ee^{-\ii\xi} + \tilde{r})(\ee^{-\ii\phi} + r_u) - d_{tu}^2\right]}.
\end{equation}
\end{widetext}
A similar expression can be obtained for the reflection coefficient $R$.
Let us note that no assumptions (aside from the neglected near-field effects) were made when deriving Eq.~\eqref{solT2} (cf.\ Subsection~\ref{ssec:cwt1}).

\subsection{Bound states in the continuum}\label{ssec:bic2}
Despite the complicated form of Eq.~\eqref{solT2}, 
an analysis similar to the one presented in Subsection~\ref{ssec:bic1} can be carried out.
First, we equate to zero both the numerator and denominator of the fraction in Eq.~\eqref{solT2}.
Then, we solve the obtained system of two equations for $\ee^{\ii \phi}$ and $\ee^{\ii\xi}$, which gives us two solutions.
The first solution has the following form:
\begin{equation}
\label{sol2a}
\ee^{\ii\phi} = \pm \frac{t}{d_{tu} d_{ru} - r_u t}, 
\;\;\;
\ee^{\ii\xi} = \pm \frac{d_{ru}}{d_{tu} t - \tilde{r} d_{ru}}.
\end{equation}
The second solution is
\begin{equation}
\label{sol2b}
\ee^{\ii\phi} = \pm \frac{1}{\sqrt{r_u \lt(r_u - d_{tu}^2 / \tilde{r}\rt)}}, 
\;\;\;
\ee^{\ii\xi} = \mp \frac{1}{\sqrt{\tilde{r} \lt(\tilde{r} - d_{tu}^2 / r_u\rt)}}.
\end{equation}

One can show that once the matrix $\mat{S}$ is unitary, the fractions in the right-hand sides of Eqs.~\eqref{sol2a} lie on the unit circle~\cite{my:Bezus:2018:pr}.
To prove this, one should note that the inverse of the unitary matrix $\mat{S}$ calculated in terms of the adjugate matrix is equal to the conjugate transpose of the matrix $\mat{S}$.
Therefore, both $\phi$ and $\xi$ in Eq.~\eqref{sol2a} are real, which gives us a BIC.
Let us note that the second solution, which is given by Eqs.~\eqref{sol2b}, does not, as a rule, describe a BIC, since the moduli of the right-hand sides of Eqs.~\eqref{sol2b} are not equal to one for an arbitrary unitary scattering matrix $\mat{S}$.
However, the BICs could accidentally result from Eqs.~\eqref{sol2b} when tuning the parameters of the grating.

Let us focus on the BICs described by Eq.~\eqref{sol2a}.
By applying the very same reasoning as in Subsection~\ref{ssec:bic1}, we obtain the following expressions for the phases:
\begin{equation}
\label{solB}
\begin{aligned}
\phi &= \pi m + \arg \frac{t}{d_{tu} d_{ru} - r_u t},\\
\xi &= \pi l + \arg \frac{d_{ru}}{d_{tu} t - \tilde{r} d_{ru}},
\end{aligned}
\end{equation}
where $m$ and $l$ are non-negative integers having the same parity~\cite{my:Bezus:2018:pr}.

Then, we solve Eqs.~\eqref{phi1} and~\eqref{xi1} to obtain the wave numbers and frequencies providing the BICs in the considered structure:
\begin{equation}
\label{bicwkx3}
k_x = \frac{\pi}{d} + \frac{d}{4\pi h^2} \lt(\phi^2 - \xi^2\rt),\;\;\;
\omega = \frac{\cc}{n} \sqrt{k_x^2+\frac{\xi^2}{h^2} }.
\end{equation}
These equations describe BICs in the region $B_1$.
Similar equations can be obtained for the region $B_2$.

\subsection{\label{ssec:sim2}Numerical simulations}

\begin{figure*}
	\centering
		\includegraphics{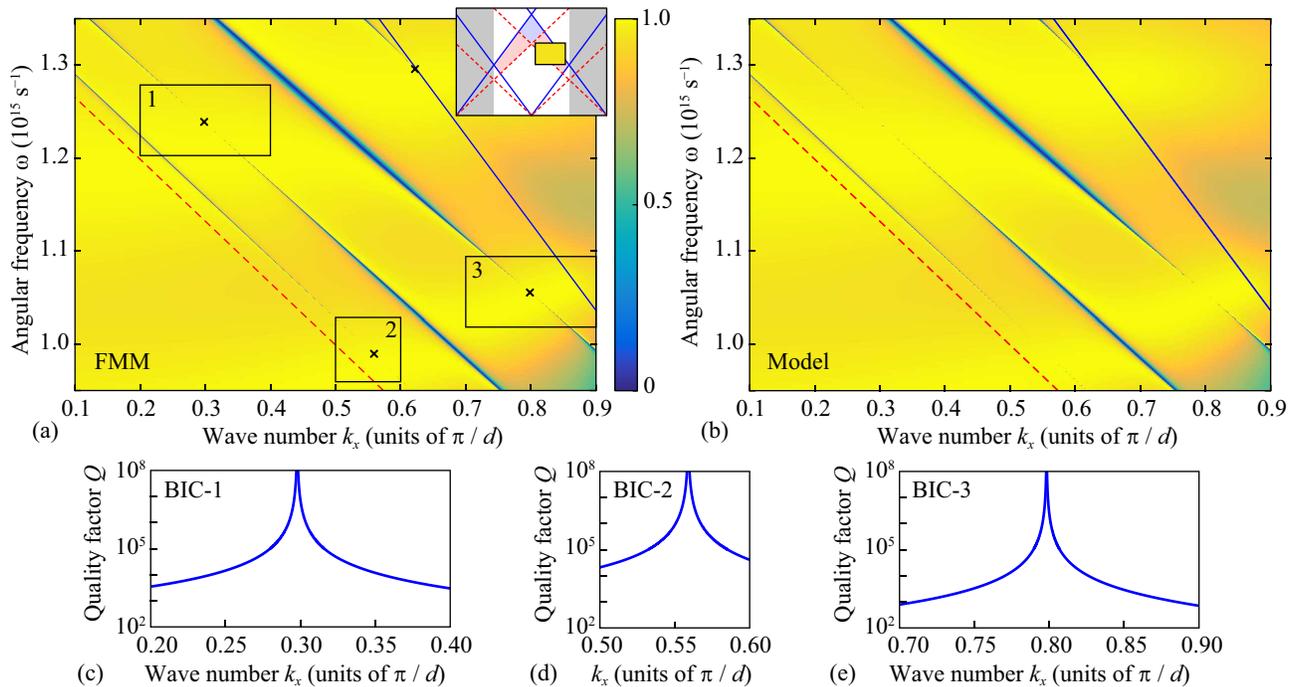}
	\caption{\label{fig:6}Rigorously calculated transmission spectrum of the considered GMRG possessing horizontal symmetry~(a). Model spectrum calculated using Eq.~\eqref{solT2} (b).
	The quality factors of the modes corresponding to the fragments marked by rectangles~(c)--(d).}
\end{figure*}

Figure~\ref{fig:6}(a) shows the $\omega$--$k_x$ transmission spectrum of a GMRG with a horizontal symmetry plane.
The spectrum was calculated using the Fourier modal method (FMM).
The parameters of the binary gratings and WG layer are the same as in Subsection~\ref{ssec:sim1}.
The spectrum exhibits~4 pronounced resonant curves, each one having a BIC.
Figure~\ref{fig:6}(b) shows the model spectrum calculated using Eq.~\eqref{solT2}.
The model spectrum perfectly agrees with the rigorously calculated one, which confirms the accuracy of the coupled-wave model of Subsection~\ref{ssec:cwt2}.

Figures~\ref{fig:6}(c)--\ref{fig:6}(e) show the quality factor of the modes. The ranges of the Q-factor plots coincide with the black rectangles in Fig.~\ref{fig:6}(a).
The divergence in the Q-factor plots confirms that the considered structure supports BICs.

In order to verify the BIC model presented in Subsection~\ref{ssec:bic2}, we calculated the BIC positions using Eqs.~\eqref{solB} and~\eqref{bicwkx3}.
The predicted $(\omega, k_x)$ points are marked in Fig.~\ref{fig:6}(a) with black crosses. The predicted BIC positions are in excellent agreement with the results of the rigorous simulations.

Let us note that the models presented in the current and in the previous sections are quite similar.
Indeed, we use either Eq.~\eqref{sol1} or Eq.~\eqref{sol2a} to describe the BIC positions.
To prove the BIC existence, we show that the right-hand sides of these equations have unit magnitude by using the consequences of the unitarity of the scattering matrix $\mat{S}$.
At the same time, the BIC formation mechanisms for the two considered cases are quite different.
Indeed, in the case of the region~A, the BICs (and quasi-BICs) emerge at the avoided crossings of the dispersion curves (see Fig.~\ref{fig:spectrum1}). 
This anticrossing indicates the strong coupling between the waveguide modes having comparable Q-factors.
However, in the region~B considered in this section, the Fabry--P{\'e}rot modes have significantly lower quality factors.
Consequently, the coupling between a waveguide mode and a Fabry--P{\'e}rot mode is weaker in this case, and no avoided crossings are present in Fig.~\ref{fig:6}. Nevertheless, both the model and the simulation results demonstrate that even this ``weak'' coupling provides the formation of the BICs.

\section{\label{sec:conclusion}Conclusion}
In this paper, we investigated bound states in the continuum (BICs) supported by lossless guided-mode resonant gratings comprising a slab waveguide and a binary grating attached to one or both
of its interfaces.
Two mechanisms behind the BIC formation were studied: the coupling of counter-propagating waveguide modes, and the coupling  of a waveguide mode with a Fabry--P{\'e}rot mode.
In both cases, the BICs (or quasi-BICs) arise due to multiwave interference of light inside the slab.
We formulated accurate coupled-wave models, which rigorously prove the BIC existence and predict their locations in the $\omega$--$k_x$ plane.
The BIC existence proof exploits the energy conservation law resulting in the unitarity of the scattering matrix of the binary grating.
The rigorous simulation results confirm high accuracy of the presented models.

The formulated models suggest that the existence of the BICs is closely connected with the symmetry properties of the gratings and, hence, with the form of their scattering matrix.
In our opinion, the models similar to the ones presented here can be developed for photonic structures possessing different symmetries.

Finally, let us outline a few particular directions for the further research.
We believe that similar models can be formulated for different parameter spaces:
instead of $(\omega,k_x)$ pair, one can consider $(\theta, h)$ or even $(k_y, h)$, with the latter corresponding to the case of ``purely conical'' diffraction (nonzero $k_y$ at $k_x = 0$).
Besides, the model of Section~\ref{sec:b} can be applied with minimal modifications when the lower grating is shifted with respect to the upper one.
We also believe that the main results of Section~\ref{sec:b} can be used to describe BICs in high-contrast gratings. In this case, we expect that Eqs.~\eqref{S2a}--\eqref{solB} will remain the same, whereas the analogue of Eq.~\eqref{bicwkx3} will be based on the dispersion relation of a 1D photonic crystal.
\\

\vspace{-2em}
\section*{Acknowledgements}
This work was funded by Russian Foundation for Basic Research 
(project nos. 18-37-20038 and 16-29-11683; coupled-wave model) 
and by Ministry of Science and Higher Education of the Russian Federation 
(State assignment to the FSRC ``Crystallography and Photonics'' RAS; numerical simulations).


\begin{thebibliography}{10}

\bibitem{NW}
J.~von Neumann und E.~Wigner,
\newblock {\"U}ber merkw{\"u}rdige diskrete Eigenwerte. {\"U}ber das Verhalten von Eigenwerten bei adiabatischen Prozessen,
\newblock {\em Phys. Zeit.} \textbf{30}, 467 (1929).

\bibitem{HsuReview}
C.\,W.~Hsu, B.~Zhen, A.\,D. Stone, J.\,D.~Joannopoulos, and M.~Solja{\v c}i{\'c}.,
\newblock {Bound states in the continuum,}
\newblock {\em Nat. Rev. Mat.} \textbf{1}, 16048 (2016).

\bibitem{FW}
H.~Friedrich and D.~Wintgen,
\newblock {Interfering resonances and bound states in the continuum,}
\newblock {\em Phys. Rev. A} \textbf{32}, 3231 (1985).




\bibitem{Marinica}
D.\,C.~Marinica, A.\,G.~Borisov, and S.\,V.~Shabanov,
\newblock {Bound states in the continuum in photonics,}
\newblock {\em Phys. Rev. Lett.} \textbf{100}, 183902 (2008).

\bibitem{Bulgakov:2018:josab}
E.\,N.~Bulgakov, D.\,N.~Maksimov, P.\,N.~Semina, and S.\,A.~Skorobogatov,
\newblock {Propagating bound states in the continuum in dielectric gratings,}
\newblock {\em J. Opt. Soc. Am. A} \textbf{35}, 1218 (2018).

\bibitem{Sadreev2}
E.\,N.~Bulgakov and A.\,F.~Sadreev,
\newblock {Bloch bound states in the radiation continuum in a periodic array of dielectric rods,}
\newblock {\em Phys. Rev. A} \textbf{90}, 053801 (2014).

\bibitem{Sadrieva}
Z.\,F.~Sadrieva and A.\,A.~Bogdanov,
\newblock {Bound state in the continuum in the one-dimensional photonic crystal slab,}
\newblock {\em J. Phys.: Conf. Ser.} \textbf{741}, 012122 (2016).

\bibitem{Yoon:2015:sr}
J.\,W.~Yoon, S.\,H~Song, and R.~Magnusson,
\newblock {Critical field enhancement of asymptotic optical bound states in the continuum,}
\newblock {\em Sci. Rep.} \textbf{5}, 18301 (2015).

\bibitem{Sadrieva2}
Z.\,F.~Sadrieva, I.\,S.~Sinev,  K.\,L.~Koshelev, A.~Samusev, I.\,V.~Iorsh, O.~Takayama, R.~Malureanu, A.\,A.~Bogdanov, and A.\,V.~Lavrinenko,
\newblock {Transition from optical bound states in the continuum to leaky resonances: Role of substrate and roughness,}
\newblock {\em ACS Photonics} \textbf{4}, 723 (2017).



\bibitem{Hsu}
C.\,W.~Hsu, B.~Zhen, J.~Lee, S.\,L.~Chua, S.\,G.~Johnson, J.\,D.~Joannopoulos, and M.~Solja{\v c}i{\'c},
\newblock {Observation of trapped light within the radiation continuum,}
\newblock {\em Nature} \textbf{499}, 188 (2013).

\bibitem{Azzam:2018:prl}
S.\,I.~Azzam, V.\,M.~Shalaev, A.\,Boltasseva, and A.\,V.~Kildishev,
\newblock {Formation of bound states in the continuum in hybrid plasmonic-photonic systems,}
\newblock {\em Phys. Rev. Lett.} \textbf{121}, 253901 (2018).

\bibitem{Shipman}
S.\,P.~Shipman and S.~Venakides,
\newblock {Resonant transmission near nonrobust periodic slab modes,}
\newblock {\em Phys. Rev. E} \textbf{71}, 026611 (2005).


\bibitem{my:Bykov:2015:pra}
D.\,A.~Bykov and L.\,L.~Doskolovich,
 \newblock {$\omega$--$k_x$ Fano line shape in photonic crystal slabs,} 
\newblock {\em Phys. Rev. A} \textbf{92}, 013845 (2015).

\bibitem{Blanchard}
C.~Blanchard, J.-P.~Hugonin, and C.~Sauvan,
\newblock {Fano resonances in photonic crystal slabs near optical bound states in the continuum,}
\newblock {\em Phys. Rev. B} \textbf{94}, 155303 (2016).

\bibitem{Bulgakov:2018:pra}
E.\,N.~Bulgakov and D.\,N.~Maksimov,
\newblock {Avoided crossings and bound states in the continuum in low-contrast dielectric gratings,}
\newblock {\em Phys. Rev. A} \textbf{98}, 053840 (2018).


\bibitem{Hsu2}
C.\,W.~Hsu, B.~Zhen, S.-L. Chua, S.\,G.~Johnson, J.\,D.~Joannopoulos, and M.~Solja{\v c}i{\'c},
\newblock {Bloch surface eigenstates within the radiation continuum,}
\newblock {\em Light Sci. Appl.} \textbf{2}, e84 (2013).

\bibitem{Sadreev1}
E.\,N.~Bulgakov and A.\,F.~Sadreev,
\newblock {Bound states in the continuum in photonic waveguides inspired by defects,}
\newblock {\em Phys. Rev. B} \textbf{78}, 075105 (2008).

\bibitem{Sadreev3}
I.\,V.~Timofeev, D.\,N.~Maksimov, and A.\,F.~Sadreev,
\newblock {Optical defect mode with tunable Q factor in a one-dimensional anisotropic photonic crystal,}
\newblock {\em Phys. Rev. B}, \textbf{97}, 024308 (2018).


\bibitem{A1}
Y.~Plotnik, O.~Peleg, F.~Dreisow, M.~Heinrich, S.~Nolte, A.~Szameit, and M.~Segev,
\newblock {Experimental observation of optical bound states in the continuum,}
\newblock {\em Phys. Rev. Lett.} \textbf{107}, 183901 (2011).


\bibitem{A2}
M.\,I.~Molina, A.\,E.~Miroshnichenko, and Y.\,S.~Kivshar,
\newblock {Surface bound states in the continuum,}
\newblock {\em Phys. Rev. Lett.} \textbf{108}, 070401 (2012).


\bibitem{A3}
S.~Weimann, Y.~Xu, R.~Keil, A.\,E.~Miroshnichenko, A.\,T{\"u}nnermann, S.~Nolte, A.\,A.~Sukhorukov, A.~Szameit, and Y.\,S.~Kivshar,
\newblock {Compact surface Fano states embedded in the continuum of waveguide arrays,}
\newblock {\em Phys. Rev. Lett.} \textbf{111}, 240403 (2013).



\bibitem{Zou}
C.\,L.~Zou, J.\,M.~Cui, F.\,W.~Sun, X.~Xiong, X.\,B.~Zou, Z.\,F.~Han, and G.\,C.~Guo,
\newblock {Guiding light through optical bound states in the continuum for ultrahigh-Q microresonators,}
\newblock {\em Laser Photon. Rev.} \textbf{9}, 114 (2015).

\bibitem{my:Bezus:2018:pr}
E.\,A.~Bezus, D.\,A.~Bykov, and L.\,L.~Doskolovich,
\newblock {Bound states in the continuum and high-Q resonances supported by a dielectric ridge on a slab waveguide,}
\newblock {\em Photon. Res.} \textbf{6}, 1084 (2018).




\bibitem{Gippius:2005:prb}
N.\,A.~Gippius, S.\,G.~Tikhodeev, and T.~Ishihara,
\newblock {Optical properties of photonic crystal slabs with an asymmetrical unit cell,} 
	\newblock {\em Phys. Rev. B} \textbf{72}, 045138 (2005). 

\bibitem{my:bykov:2013:jlt}
D.\,A.~Bykov and L.\,L.~Doskolovich,
\newblock {Numerical methods for calculating poles of the scattering matrix with applications in grating theory,} 
	\newblock {\em J. Lightw. Technol.} \textbf{31}, 793 (2013).


\bibitem{Moharam:1995:josaa}
M.\,G.~Moharam, E.\,B.~Grann, D.\,A. Pommet, and T.\,K.~Gaylord,
\newblock {Formulation for stable and efficient implementation of the rigorous coupled-wave analysis of binary gratings,} 
\newblock {\em J. Opt. Soc. Am. A} \textbf{12}, 1068 (1995).


\bibitem{Li:1996:josaa2}
L.~Li,
\newblock {Formulation and comparison of two recursive matrix algorithms for modeling layered diffraction gratings,} 
\newblock {\em J. Opt. Soc. Am. A} \textbf{13}, 1024 (1996).

\end{thebibliography}
\end{document}